\documentclass[twocolumn,showpacs,preprintnumbers]{revtex4}
\usepackage{dcolumn}
\usepackage{bm}
\usepackage{amsmath}
\usepackage{graphicx}
\usepackage{mathrsfs}
\usepackage{amssymb}
\usepackage{amsfonts}

\begin{document}
\title{Parity breaking and scaling behavior in light-matter interaction}
\author{T. Liu,$^{1,2}$\footnote{E-mail:liutao849@163.com}
M. Feng,$^{1}$\footnote{E-mail:mangfeng@wipm.ac.cn} W. L. Yang,$^{1}$ J. H. Zou,$^{1}$
L. Li,$^{2}$ Y. X. Fan,$^{2}$ and K. L. Wang$^{3}$}
\affiliation{$^{1}$  State Key Laboratory of Magnetic Resonance and
Atomic and Molecular Physics, Wuhan Institute of Physics and
Mathematics, Chinese Academy of Sciences, Wuhan, 430071, China \\
$^{2}$ The School of Science, Southwest University of Science and Technology, Mianyang 621010, China \\
$^{3}$ The Department of Modern Physics, University of Science and
Technology of China, Hefei 230026, China}

\begin{abstract}
The light-matter interaction described by Rabi model and
Jaynes-Cummings (JC) model is investigated by parity breaking as
well as the scaling behavior of ground-state population-inversion
expectation. We show that the parity breaking leads to different
scaling behaviors in the two models, where the Rabi model
demonstrates scaling invariance, but the JC model behaves in
cusp-like way. Our study helps further understanding rotating-wave
approximation and could present more subtle physics than any other
characteristic parameter for the difference between the two models.
More importantly, our results could be straightforwardly applied to
the understanding of quantum phase transitions in spin-boson model.
Furthermore, the scaling behavior
is observable using currently available techniques in light-matter
interaction.
\end{abstract}
\maketitle

The Rabi model \cite{rabi} presents an important prototype for the
interaction between a single two-level system (e.g., a spin) and a
quantum bosonic field, which has been widely applied to almost every
subfield of physics, e.g., the cavity quantum electrodynamics (QED)
and the exciton-photon interaction. Under the rotating-wave
approximation (RWA), the Rabi model is reduced to the
Jaynes-Cummings (JC) model \cite{jc} in which counter-rotating
effects are neglected by the assumption of large detuning and small
Rabi frequency. It is generally believed that the RWA works worse
and worse with the increase of the Rabi frequency \cite{sk}. So, in
the case of strong spin-field coupling, the Rabi model, rather than
the JC model, is required. The differences between the two models
have been studied by solving the eigenenergies \cite{non-rwa,liu1}, the
dynamics \cite{non-rwa-1,dyna}, the Berry phase \cite{berry} and so
on.

Going beyond the interaction between a single spin-1/2 and a single
quantum mode, the Rabi model has been extended to a big-spin
(S$>$1/2) system \cite{muiti} or many spin-1/2 experiencing a single
quantum mode, the latter of which is called Dicke model
\cite{dicke}. It has been shown that the RWA introduced in the
Hamiltonian of the Dicke model brings about completely different
phenomena from the non-RWA case in the quantum phase transition
\cite{buzek, zhang}. Besides, we may also consider a single spin-1/2
interacting with a multi-mode quantum bosonic field, called
spin-boson model \cite{leggett, weiss}, to describe the dissipation
of a single spin under the bosonic bath. The spin-boson models with
and without the RWA demonstrate different behaviors in the spin
dissipation \cite{hur}.

We focus  in the present work on the scaling behaviors in the Rabi
and JC models, which could present us more subtle physics and more
evident difference than the solutions of eigenenergies, geometric
phases and dynamical properties. The different scaling behaviors are
relevant to different symmetries and can be understood by parity
breaking. Specifically, we show that the scaling invariance exists
only in the Rabi model. In contrast, the scaling behavior in JC
model behaves much differently with the change of some
characteristic parameters, such as the detuning. So the difference
between the two models turns to be more evident in the observation
of the scaling behavior because of the significant change of the
symmetries in the two models due to the introduction of the local
bias field. By extending the idea to the multi-mode case, we may
discover the deeper physics hidden in the quantum phase transitions
in spin-boson model. More importantly, these scaling behaviors are
strongly relevant to the dynamics of the spin, which could be
observed in some experimentally available systems, such as the
circuit QED, the trapped ion and the nanomechanical systems.

We get started from the following Hamiltonian ($\hbar =1$ throughout
the work) \cite{explain},
\begin{equation}
H_{sb}=-\frac{\Delta }{2}\sigma _{x}+\frac{\varepsilon }{2}\sigma
_{z}+\omega a^{\dagger }a+\lambda (a+a^{\dagger})\sigma_{z}
\label{1},
\end{equation}
where $\Delta$ and $\varepsilon$ are the tunneling and the local
bias field, respectively, $\omega$ and $a^{\dagger}$ ($a$) are
frequency and the creation (annihilation) operator of the
single-mode bosonic field, and $\lambda$ is the Rabi frequency.
$\sigma_{z,x}$ are the usual Pauli operators for the spin-1/2 and
$\sigma_{x}=\sigma_{+}+\sigma_{-}$ with $\sigma_{\pm}=(\sigma_{x}\pm
i\sigma_{y})/2$. Compared to the standard form of the Rabi model,
Eq. (1) owns an additional term, i.e., the local bias, which brings in
new and more interesting physics. As discussed later,
Eq. (1) connects directly to the spin-boson model and to other currently
experimentally achieved systems.

Eq. (1) can be diagonalized by displaced coherent states. The
eigenfunction of $H_{sb}$ has the form \cite {berry,liu1}
\[
|\tilde{\Psi}\rangle =\sum_{n}\left(
\begin{array}{l}
~~~~~~~~~~~c_{n}|n\rangle _{A} \\
(-1)^{n+1}d_{n}|n\rangle _{B}
\end{array}
\right),
\]
where $c_{n}$ and $d_{n}$ are coefficients to be determined later,
$|n\rangle _{A}=\frac{e^{-q^{2}/2}}{\sqrt{n!}}(a^{\dagger
}+q)^{n}e^{-qa^{\dagger}}|0\rangle$ and $|n\rangle _{B}
=\frac{e^{-q^{2}/2}}{\sqrt{n!}}(a^{\dagger
}-q)^{n}e^{qa^{\dagger}}|0\rangle$ are the displaced coherent states
with the displacement variable $q=\lambda /\omega $. The complete
eigensolution of H$_{sb}$ can be obtained from the
Schr$\stackrel{..}{o}$dinger equation (See Appendix A)
\cite{explain1}. We check the ground-state population-inversion
$\langle \sigma_{z}\rangle$, which is,
\begin{equation}
\langle \sigma _{z}\rangle
=(c_{0}^{-})^{2}-(d_{0}^{-})^{2}=\frac{-\kappa }{\sqrt{\kappa
^{2}+e^{-4\beta }}},  \label{2}
\end{equation}
where $\beta =q^{2}$, $\kappa =\varepsilon /\Delta$, and $c_{0}^{-}$
and $d_{0}^{-}$ are defined in Appendix A. In contrast to the
perturbation solutions published previously \cite{pertur}, we obtain
Eq. (2) by considering the zeroth-order elements in the
diagolization of the matrix of Eq. (1) \cite {liu1}. In the present
case of very small $\Delta$, this consideration is a very good
approximation of the complete solution to Eq. (1) \cite {explain2}.

We first perform the second derivative of $\langle
\sigma_{z}\rangle$ with respect to $\beta$, which yields a
reflection point $\beta_{c}=-\ln (2\kappa^{2})/4$ and thereby Eq.
(2) is rewritten as
\begin{equation}
\langle \sigma _{z}\rangle =\frac{-\kappa }{\sqrt{\kappa
^{2}+e^{\beta ^{^{\prime }}\ln (2\kappa ^{2})}}},  \label{3}
\end{equation}
under the scaling transformation $\beta^{\prime}=\beta /\beta_{c}$.
From the definition of $\beta_{c}$, we require $\kappa\neq 0$, i.e.,
$\epsilon\neq 0$ in the present calculation. For a fixed value of
$\kappa$, the population-inversion $\langle\sigma _{z}\rangle $ in
Eq. (3) is only relevant to $\beta^{^{\prime}}$, rather than to
other characteristic parameters (See Fig. 1(a)). So $\beta_{c}$ can
be regarded as a scale of the Rabi model. In addition, if we set
$\beta ^{^{\prime}}=1$, the population-inversion
$\langle\sigma_{z}\rangle$ turns to be a constant $-1/\sqrt{3}$,
implying a fixed crossing point with variation of $\beta^{^{\prime
}}$. It is more interesting to demonstrate the scaling behavior of
the population-inversion $\langle \sigma _{z}\rangle $ with a
displaced scaling $\beta ^{^{^{\prime\prime }}}=(\beta -\beta
_{c})/\sqrt{27}$. Since $\langle \sigma _{z}(\beta
^{^{^{\prime\prime}}})\rangle =-1/\sqrt{1+2e^{-12\sqrt{3}\beta
^{^{^{\prime \prime}}}}}$, which is independent of $\kappa$ under
the scaling transformation, the population-inversion $\langle \sigma
_{z}\rangle $ with respect to $\beta^{^{^{\prime\prime}}}$, remains
unchanged for different parameters $\kappa$, as shown in Fig. 1(b).
\begin{figure}[tbph]
\centering
\includegraphics[width=3.4 in]{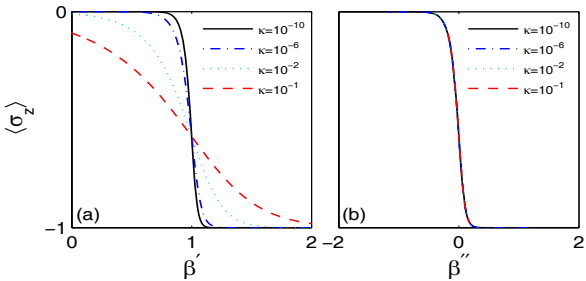}
\par
\hspace{2cm} \vspace{-1.0cm} \caption{(color online) Scaling
behavior of the ground-state population-inversion. (a) As a function
of $\beta'$ which has a fixed point; (b) As a function of
$\beta^{\prime\prime}$ which remains unchanged. } \label{fig1}
\end{figure}

It has been generally considered that the scaling invariance is
relevant to the critical points of quantum phase transition in
spin-boson model \cite {hur}. Despite only for the interaction
between a single spin and a single mode of the quantized field,
the scaling behavior we show here can discover the physical insight
in the fundamental element of the spin-boson model, which is resulted
from the parity breaking regarding the parity operator $\Pi
=\sigma_{x}e^{i\pi a^{\dagger}a}$. If we denote the case of
$\epsilon =0$ in Eq. (1) by $H_{sb}^{^{\prime }}$, we have
$[H_{sb}^{^{\prime }},\Pi]=0$, with the ground state of their common
eigenfunction to be $|\psi_{0}^{-}\rangle =-\frac{1}{\sqrt{2}}\left(
\begin{array}{l}
|0\rangle _{A} \\
|0\rangle _{B}
\end{array}
\right) $satisfying $\Pi|\psi_{0}^{-}\rangle =|\psi_{0}^{-}\rangle
$, i.e., an even parity state of $\Pi$. The even parity breaks down
in the variation from $\varepsilon =0$ to $\epsilon\neq 0$ because
the Hamiltonian $H_{sb}$ never commutes with the parity operator
$\Pi$, i.e., $[H_{sb}, \Pi]\neq 0$. This parity-breaking case
plotted in Fig. 1(a) (except the point $\beta'=0$) demonstrates the
abrupt variation of $\langle\sigma_{z}\rangle$ from 0 to -1 at
$\kappa$ approaching 0, corresponding to the translation from spin
non-localization (i.e., superposition of the spin-up and spin-down)
to spin localization (i.e., the spin down). This peculiar scaling
behavior is very analogous to the critical point behavior in quantum
phase transition of the spin-boson model (See Fig. 3 in \cite{hur}).
More interestingly, this sharp scaling variation induced by the
parity breaking could appear for larger $\kappa$ (e.g.,
$\kappa=$0.1) under a scaling displacement (Fig. 1(b)).

\begin{figure}[tbph]
\centering
\includegraphics[width=3.4 in]{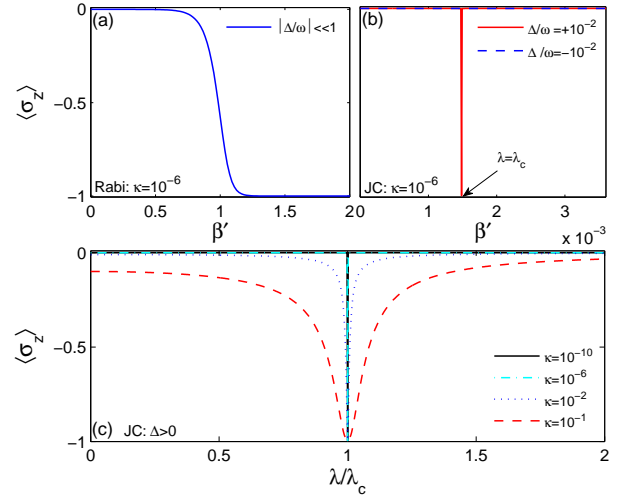}
\par
\caption{(color online) (a) The ground-state population-inversion
for $H_{sb}$ as a function of $\beta^{\prime }$ with $\kappa
=10^{-6}$. (b) The ground-state population-inversion for $H_{jc}$ as
a function of $\beta^{\prime}$ for $\Delta >0$ (the red solid line
with cusp-like scaling behavior) and for negative $\Delta$ (the blue
dashed line with no scaling behavior) with $\kappa=10^{-6}$. (c) The
cusp-like scaling behavior of the ground-state population-inversion
for $H_{jc}$ as a function of $\lambda /\lambda_{c}$ with $\Delta
>0$ for $\kappa =10^{-10}$ (the black solid line), $\kappa=10^{-6}$
(the blue dashed-dotted line), $\kappa =10^{-2}$ (the pink dotted
line), and $\kappa =10^{-1}$ (the red dashed line). For convenience
of comparison between (b) and (c), we label $\lambda_{c}$ in (b). } \label{fig2}
\end{figure}

To better understand the parity breaking in the Rabi model, we may
consider the same treatment for the JC model. In the case of
$\epsilon=0$, the ground-state of the JC model is usually of certain
parity, and experiences energy level crossing. In the case of
$\epsilon\neq 0$, however, the parity breaks down and level crossing
turns to level avoided-crossing (See Appendix B). But different from
the scaling invariance in the Rabi model, the variation of
$\langle\sigma_{z}\rangle$ with respect to $\beta'$ behaves with a
cusp in the JC model (See Fig. 2(a) and 2(b) for comparison between
the two models). It implies that the spin is always in
non-localization except the rare cases near the critical point
$\lambda_{c}$ for localization. For $\kappa=10^{-6}$, we have
$\beta_{c}=6.73$, corresponding to $\lambda=2.59\omega$ in the Rabi
model and to $\lambda=0.09\omega$ in the JC model. These are
reasonable values for the parameters in the two models. For a more
clarified demonstration, we plot in Fig. 2(c)
$\langle\sigma_{z}\rangle$ versus $\lambda/\lambda_{c}$, where
$\lambda_{c}=\sqrt{\omega\Delta}$ takes values much smaller than
$\beta_{c}$. So we could have a zooming-in picture for the variation
of $\langle\sigma_{z}\rangle$, which shows strong relevance to
$\kappa$. The cusp-like behavior is evidently induced by the parity
breaking (i.e., for $\kappa$ very close to zero).

The results above remind us of the role the counter-rotating terms
playing in the light-matter interaction. For the parity operator
$\Pi$ with respect to H$_{sb}$, we have $\Pi'=U^{\dagger}\Pi
U=-\sigma_{z}e^{i\pi a^{\dagger}a}$ for H$_{jc}$ where $U$ is
defined in \cite{explain}. If we neglect the terms regarding
$\epsilon$, we have $[\Pi, H'_{sb}]=0$ and $[\Pi', H_{rjc}]=0$. But
this does not mean that the counter-rotating terms have no extra
influence on the symmetry in the Rabi model compared to the JC
model. First, in the case of $\epsilon=0$, the ground-state of the
JC model (for $\Delta>$0) experiences a translation from the even
parity to the odd in the increase of the coupling $\lambda$, but the
ground-state of the Rabi model only stays in the even parity. This
could be understood physically from whether the energy level
crossing happens or not. But the deeper physics is the different
parity behaviors in both models, which yield different dynamics
\cite{dyna}. Second, for $\epsilon\neq 0$, only energy
avoided-crossing occurs in either of the models (See Appendix B). We
may discern the two models by scaling behaviors around the critical
point of the parity breaking, which presents the difference between
the two models more evidently than other quantities \cite
{non-rwa,liu1,berry}. This idea can be straightforwardly extended to
the spin-boson model: The difference between the quantum phase
transitions with and without RWA should be relevant to different
critical behaviors around the parity breaking points.

Eq. (1) can be generated in various systems. Since
$\langle\sigma_{z}\rangle$ is observable experimentally, we may
demonstrate the scaling behaviors using currently available
techniques. We first check the circuit QED system with a
superconducting qubit strongly coupled to a microwave resonator mode
via external driving, which can be described by
\cite{cir1,cir2,ball} $H_{qed}=\frac{\omega_{q}}{2}\sigma_{z} +
\omega_{b} b^{\dagger }b + G(b\sigma_{+} + b^{\dagger}\sigma_{-}) +
\Omega_{1}(e^{i\omega_{1}t}\sigma_{-}+e^{-i\omega_{1}t}\sigma _{+})
-\Omega_{2}(e^{i\omega_{2}t}\sigma_{-}+e^{-i\omega_{2}t}\sigma_{+})$,
where $\omega_{q}$ and $\omega_{b}$ are, respectively, the qubit and
microwave photon frequencies with $G$ the qubit-photon coupling
strength. $b$ ($b^{\dagger}$) stands for the annihilation (creation)
operator of the microwave photon. $\sigma_{z, \pm}$ are usual Pauli
operators for the superconducting qubit. $\Omega_{j}$ and
$\omega_{j}$ are the amplitude and frequency of the $j$th driving
field ($j=$1, 2). We assume $\Omega_{1}\gg \Omega_{2}$, $G$ and in
the rotating frame first with the driving field $\omega_{1}$ and
then with a large frequency $\Omega_{3}$ comparable to $\Omega_{1}$,
we may obtain an effective Hamiltonian by setting $\Omega
_{3}=(\omega _{1}-\omega_{2})/2$ and neglecting fast oscillating
terms, $H_{eff}=\frac {\Omega_{2}}{2}\sigma_{z}+ \frac
{(\Omega_{1}-\Omega_{3})}{2}\sigma_{x} +
(\omega_{b}-\omega_{1})b^{\dagger}b + \frac{G}{2}(b +
b^{\dagger})\sigma_{x}$, which can be used to simulate the Rabi
model and JC model by tuning the characteristic parameters. For the
Rabi model, to meet the condition $\Omega _{2}\ll (\omega
_{b}-\omega_{1})\sim G/2$, we can adopt following experimental
parameters as $\omega _{q}=2\pi\times 6.02$ GHz, $\omega _{b}=2\pi
\times 6.02$ GHz, $\omega _{1}=2\pi \times 6$ GHz, $\omega _{2}=2\pi
\times 4$ GHz, and a tunable $G>2\pi \times 40$ MHz. Besides, we may
assume the driving fields with the amplitudes $\Omega
_{3}=2\pi\times 1$ GHz and $\Omega_{2}=2\pi\times 0.4$ MHz.
$(\Omega_{1}-\Omega_{3})$ can change from zero to $2\pi\times$20
kHz, which are realistic values using state-of-the-art circuit-QED
technology \cite{super1,super2}. While for JC model, the result can
be obtained by reducing the coupling strength to $G < 2\pi\times 4$
MHz and remaining other parameters unchanged.

Similarly, Eq. (1) can also be implemented by a diamond
nitrogen-vacancy center (i.e., a single spin) coupled by a
nanomechanical resonator \cite {rabl}, in which the strong coupling
due to magnetic field gradient from the tip of the resonator can be
of the order of hundreds of kHz \cite {tip}. Under a driving on the
spin, we may have in the rotating frame with respect to the driving
frequency $H_{nv}=(\Omega'/2)\sigma_{x}+(\Delta'/2)\sigma_{z}+
\omega_{m}a^{\dagger}a+\lambda'(a^{\dagger}+a)\sigma_{z}$ with
$\Delta'$ the detuning of the two-level resonance frequency to the
the driving frequency, $a^{\dagger} (a)$ for the vibration of the
resonator with frequency $\omega_{m}$, and $\lambda'$ and $\Omega'$
being, respectively, the magnetic coupling and the driving strength.
By tuning the driving strength and the driving frequency, we may
simulate the Rabi model (with $\Omega'\ll\omega_{m}\sim\lambda'$)
and the JC model (with $\omega_{m}\gg\lambda'$) from
$H_{nv}$ as our will.

Simply speaking, once the strong coupling between the spin and the
quantum field is achieved, the two models can be generated from Eq.
(1) by balancing the vibrational frequency and the spin-field
coupling. In this sense, we may also consider a single trapped
ultracold ion experiencing irradiation of traveling lasers. Under
some unitary transformations \cite {liu0}, we have $H_{ion}=
-(\tilde{\Omega}/2)\sigma_{x}+\tilde{\epsilon}\sigma_{z}+
\tilde{\nu}a^{\dagger}a+\tilde{g}(a^{\dagger}+ a)\sigma_{z}$
\cite{explain3}, where $\sigma_{x,z}$ are the usual Pauli operators
for two internal levels of the ion, $a^{\dagger} (a)$ is for the
vibration of the ion, $\tilde{\Omega}$ and $\tilde{\nu}$ are the
Rabi frequency and the trap frequency, $\tilde{\epsilon}$ is
relevant to detuning and $\tilde{g}=\tilde{\nu}\eta/2$ with the
Lamb-Dicke parameter $\eta$. Since $H_{ion}$ works for any value of
$\eta$, we may have $\tilde{\Omega}\ll\tilde{\nu}\approx\tilde{g}$
by using weak laser irradiation and deep trap potential, which is
for the Rabi model, and have $\tilde{\nu}\gg\tilde{g}$ by restricting the
ion within Lamb-Dicke regime, corresponding to the JC model. The
typical trap frequency $\tilde{\nu}$ varies from hundreds of kHz to
several MHz and Rabi frequency $\tilde{\Omega}$ is adjustable by laser
irradiation (up to hundreds of kHz) \cite {blattm}.

In conclusion, we have investigated the scaling behaviors of
ground-state population-inversion in light-matter interaction, which
is relevant to the parity breaking. The different scaling behaviors
can be used to not only identify the difference between the Rabi
model and the JC model, but also demonstrate the change of the
symmetry in the system. We have also discussed the experimental
feasibility of our study in different systems using currently
available technology. The present idea could be straightforwardly
extended to the study of spin-boson model for possible quantum phase
transitions induced by the parity breaking.

This work is supported by NFRPC (Grant No. 2012CB922102) and by NNSFC
(Grants No. 11274352, No. 11147153, No. 11004226,
and No. 11147153).

\section*{APPENDIX}

\textbf{A. Our eigensolution to the Rabi model}\\
\\
Taking $|\tilde{\Psi}\rangle$ into the Schr$\stackrel{..}{o}$dinger
equation of H$_{sb}$, we obtain a set of equations $
\lbrack \omega (m-q^{2})+\frac{\epsilon }{2}]c_{m}+\frac{\Delta }{2}
\sum_{n} d_{n}D_{m,n}=Ec_{m}$ and $\lbrack \omega (m-q^{2})-
\frac{\epsilon }{2}]d_{m}+\frac{\Delta }{2}
\sum_{n} c_{n}D_{m,n}=Ed_{m}$, with $D_{m,n}=e^{-2q^{2}}\sum_{k=0}^{\min [m,n]}(-1)^{-k}
\frac{\sqrt{m!n!}(2q)^{m+n-2k}}{(m-k)!(n-k)!k!}$ \cite{berry,
liu0,liu1}. In our case with the condition $\Delta/\omega \ll 1$,
the terms of $D_{m,n}$ with $m\neq n$ play negligible roles in the
equations compared to other terms with $m=n$. So the equations above
can be simplified to the case remaining the terms of $m=n$, which
yields following analytical solutions, that is, the eigenenergies
$E_{m}^{\pm}=\omega
(m-q^{2})\pm\sqrt{\epsilon^{2}+\Delta^{2}D_{m,m}^{2}}/2$, and the
coefficients $c_{m}^{\pm }=\mu_{m}^{\pm }/\sqrt{1+(\mu_{m}^{\pm
})^{2}}$ and $d_{m}^{\pm }=1/\sqrt{1+(\mu_{m}^{\pm})^{2}}$ with
$\mu_{m}^{\pm}=[\epsilon\pm\sqrt{\epsilon^{2}+\Delta
^{2}D_{m,m}^{2}}]/(\Delta D_{m,m})$. The eigenfunction of the
ground-state is $|\psi_{0}^{-}\rangle =-\left(\begin{array}{l}
|0\rangle_{A} \\
|0\rangle_{B}
\end{array}
\right)/\sqrt{2}$ since
$E_{0}^{-}$ is smaller than $E_{0}^{+}$.
\\

\textbf{B. Avoided-crossing around the critical points of parity breaking}\\
\\
\begin{figure}[tbph]
\centering
\includegraphics[width=3.4 in]{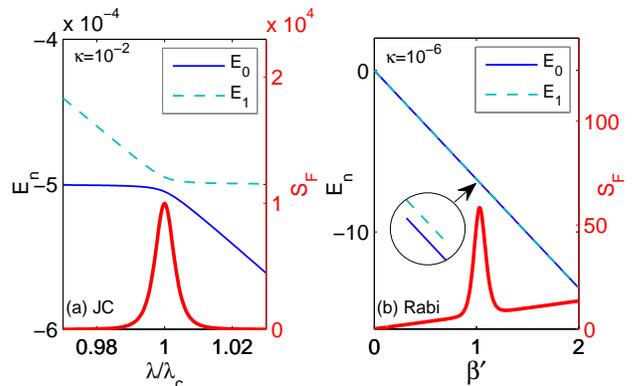}
\par
\hspace{2cm} \vspace{-1.0cm} \caption{(color online) Eiegenenergies
and fidelity susceptibility S$_{F}$. (a) for JC model with the
additional term $\epsilon/2$, no level crossing appears and sudden
change in fidelity susceptibility occurs around
$\lambda=\lambda_{c}$; (b) for Rabi model with the additional term
$\epsilon/2$, no level crossing appear and sudden change in fidelity
susceptibility occurs around $\beta'=1$. } \label{fig3}
\end{figure}

In JC model, there are level crossings in the ground state with the
excited states in different parities. So a level crossing means the
change of the parity in the ground state. In contrast, there are
only avoided-crossings of levels in Rabi model, which yield
unchanged parity. So the RWA effect can be
reflected from the parity change. In our present work,
however, due to the introduction of the additional term regarding
$\epsilon/2$, no level crossing exists anymore in the JC model, but
only avoided-crossings, as shown in Fig. 3. So we may consider other
characteristic properties, such as scaling behavior, for
understanding the effect of the RWA. Our results show that the
scaling behavior can present more evident difference than any other
characteristic parameters used previously. More interestingly, we
see in Fig. 3 the sudden change occurring in the fidelity
susceptibility (S$_{F}$) \cite{fs} around the critical point of the
parity breaking, which also happens in the spin-boson model around
the critical point of quantum phase transition \cite{sf1}. This
implies that the deeper physics for quantum phase transition in the
multi-mode case might be also due to the parity breaking. The
fidelity susceptibility is defined by $
S_{F}(\lambda_{1})=\sum_{n\neq 0} \frac {|\langle\phi_{n}(\lambda_{1})|H_{1}
|\phi_{0}(\lambda_{1})\rangle|^{2}}{[E_{n}(\lambda_{1})-E_{0}({\lambda_{1}})]^{2}}$,
where $H=H_{0}+\lambda_{1}H_{1}$ with $\lambda_{1}$ the order parameter,
$H$ is the hamiltonian under our consideration. $\phi_{k}$ and
$E_{k}$ are k$th$ eigenstate and the corresponding eigenenergy. In
our treatment, we take spin-field interaction as $H_{1}$.


\begin{references}
\bibitem{explain} The Hamiltonian is a simplified form of the spin-boson model with
the single mode of the bosonic field, which is physically equivalent
to the Rabi model with an additional local bias. Using a unitary
transformation $U=\frac{1}{ \sqrt{2}}(\begin{array}{ll}
1 & 1 \\
-1 & 1
\end{array}
)$, we may transform $H_{sb}$ to
$H_{rb}=U^{\dagger}H_{sb}U=\frac{\Delta }{2}\sigma
_{z}+\frac{\epsilon }{2}\sigma _{x}+\omega a^{\dagger }a+\lambda
(a+a^{\dagger})\sigma_{x}$,  which is the standard Hamiltonian of
the Rabi model, in addition to a tunneling term. The Rabi model
includes both the rotating terms ($a\sigma _{+}$, $a^{\dagger
}\sigma _{-}$) and the counter-rotating terms ($a\sigma $,
$a^{\dagger }\sigma _{+}$). If the counter-rotating terms are
dropped, the model is reduced to a general JC model
$H_{jc}=\frac{\Delta }{2}\sigma_{z}+\frac{ \epsilon }{2}\sigma
_{x}+\omega a^{\dagger}a + \lambda (a\sigma _{+}+a^{\dagger }\sigma
_{-})$ with a tunneling term. In our calculation, we present the
results regarding Rabi model from H$_{sb}$ in Eq. (1), and obtain
the results regarding JC model from $H_{jc}$ following the unitary
rotation $U^{\dagger}$.
\bibitem{explain1} We consider the solution of the Schr{\"o}dinger equation in the
case of $|\Delta|/\omega\ll 1$, which is based on following reasons.
First, it simplifies the solution and leads to analytical
expressions of eigenenergy and eigenfunction. This condition is also
experimentally acceptable. Second, The small value of $\Delta$ can
somewhat reflect the physics related to the parity operator
$P=\sigma_{z}$, which commutes with H$_{sb}$ in the case of
$\Delta=0$. So besides the parity breaking regarding $\Pi$ under our
consideration, the model also experiences another parity breaking if
we change $\Delta$ from zero to non-zero. But we only focus on the
parity breaking regarding $\Pi$ in the present work because we are
considering the bias field $\frac{\varepsilon}{2}\sigma_{z}$ acting
on a Rabi or JC model. The physics related to the parity operator
$P$ will be discussed in details in a separate paper.
\bibitem{explain2} We have also made numerical calculation in saturation
for Fig. 1, which presents the same results as from Eq. (2).
\bibitem{explain3} In comparison to Eq. (3) in \cite {liu0}, we omit the constant term and make up the trap frequency
$\tilde{\nu}$.
\end{references}
\end{document}